\documentclass[a4paper]{article}

\begin{document}

\title {\bf Relativistic Quantum Newton's Law and
Photon Trajectories}

\author{T.~Djama\thanks{E-mail:
{\tt djam\_touf@yahoo.fr}}}

\maketitle
\begin{center}Universit\'e des Sciences et de la
Technologie
Houari Boum\'edienne, \\Alger, Algeria \\ 
\vspace*{0.5cm}
Mail address:14, rue Si El Hou\`es,
B{\'e}ja{\"\i}a, Algeria
\end{center}

\begin{abstract}
\noindent
Using the relativistic quantum Hamilton-Jacobi
equation within
the framework of the equivalence postulate, and
grounding oneself on both relativistic and quantum Lagrangians, we
construct
a Lagrangian of relativistic quantum system in one
dimension
and derive a third order equation of motion
representing
a first integral of the relativistic quantum Newton's
law.
Then, we investigate the free particle case and
establish
the photon's trajectories.
\end{abstract}

\vskip\baselineskip



\newpage
%
\vskip0.5\baselineskip
\noindent
{\bf 1\ Introduction}
\vskip0.5\baselineskip
%

Deriving quantum mechanics from an equivalence
postulate,
Faraggi and Matone showed that the Schr\"odinger wave
function
must have the form \cite{FM1,FM2,FM3,FM4}
\begin{equation}
\phi(x)=\left({\partial S_0 \over \partial x}
\right)^{-{1 \over 2}}\;
\left[\alpha\exp \left ({i \over \hbar   } S_0
\right)+
\beta\exp \left(-{i \over \hbar}S_0\right)\right]\; ,
\end{equation}
where $\alpha$ and $\beta$ are complex constants and
$S_0$ a real function representing the quantum
reduced action.
They established that the conjugate momentum given by
\begin{equation}
P={\partial S_0 \over \partial x}
\end{equation}
never vanishes for bound and unbound states making
possible
a dynamical approach of the quantum motion of
particles.
This conjugate momentum is always real even in
classically
forbidden regions. They showed also, within the
framework
of differential geometry \cite{FM1,FM2,FM3,FM4},
that the quantum stationary Hamilton-Jacobi equation
(QSHJE)
which leads to the Shr\"odinger equation is
\begin{equation}
{1 \over 2m_0}\left({\partial S_0 \over \partial
x}\right)^2+
V(x)-E={\hbar^2 \over 4m_0}
\left[{3 \over 2}
\left({\partial S_0 \over \partial x}\right)^{-2}
\left({\partial^2 S_0 \over \partial x^2}\right)^2-
\left({\partial S_0 \over \partial x}\right)^{-1}
\left({\partial^3 S_0 \over \partial x^3}\right)
\right]
\end{equation}
where $V(x)$ is the potential and $E$ the energy. The
solution of Eq. (3) investigated by Floyd
\cite{Floyd1, Floyd2, Floyd3, Floyd4} and
Faraggi and Matone \cite{FM1,FM2,FM3,FM4} is given in
Ref. [9] as
\begin{equation}
S_0=\hbar \arctan\left(a{\theta \over \phi}+b\right)\;
,
\end{equation}
where $a$ and $b$ are real constants. $\theta$
and $\phi$ are two real independent solutions of the
Schr\"odinger equation. Taking advantage on these
results,
Bouda and Djama have recently introduced a quantum
Lagrangian
\begin{equation}
L(x,\dot{x},\mu,\nu)={1\over 2} m {\dot{x}}^2
f(x,\mu,\nu)-V(x) \; ,
\end{equation}
from which they derived the quantum law of motion 
\cite{BD1}.
They stated that the conjugate momentum of the non
relativistic
and spin-less particle is written as
\begin {equation}
{\partial S_0 \over \partial x}=
{2(E-V) \over \dot{x}} \; .
\end {equation}
From this last equation, they derived the first
integral of the
quantum Newton's law (FIQNL).
\begin {eqnarray}
(E-V)^4-{m{\dot{x}}^2 \over 2}(E-V)^3+{{\hbar}^2 \over
8}
{\left[{3 \over 2}
{\left({\ddot{x} \over
\dot{x}}\right)}^2-{\dot{\ddot{x}} \over \dot{x}}
\right]} (E-V)^2\hskip15mm&& \nonumber\\
-{{\hbar}^2\over 8}{\left[{\dot{x}}^2
{d^2 V\over dx^2}+{\ddot{x}}{dV \over dx}
 \right]}(E-V)-{3{\hbar}^2\over 16}{\left[\dot{x}
{dV \over dx}\right]^2}=0 \; ,
\end {eqnarray}
which goes at the classical limit $(\hbar \to 0)$ to the
classical conservation equation
\begin {equation}
{m\; \dot{x}^2 \over 2}+V(x)=E \; .
\end {equation}
Bouda and Djama have also
plotted some trajectories of the particle for several
potentials
\cite{BD2} .

The construction of the Lagrangian (5) and the
establishment of
Eqs. (6) and (7) are important steps to build a
deterministic
theory which restores the existence of trajectories
\cite{BD1,BD2} .
Nevertheless, such as formalism cannot approach both
relativistic
velocities cases, and more than one dimension motions.
The aim of
this paper is to generalize the dynamical formalism
that
we have recalled above \cite{BD1} into the one
dimensional relativistic velocities cases.
In this purpose let us recall the finding of Faraggi,
Matone and
Bertoldi concerning the relativistic quantum systems.
They stated
that the relativistic quantum wave function is given
by Eq. (1),
where $S_0$ defines the relativistic quantum reduced
action, and wrote
the relativistic quantum stationary Hamilton-Jacobi
equation (RQSHJ)
as \cite{FM3,FM4}
\begin{eqnarray}
{1 \over 2m_0}\left({\partial S_0 \over \partial
x}\right)^2-
{\hbar^2 \over 4m_0}\left[{3 \over 2}
\left({\partial S_0 \over \partial x}\right)^{-2}
\left({\partial^2 S_0 \over \partial
x^2}\right)^2-\right.
\hskip35mm&& \nonumber\\
\left.\left({\partial S_0 \over \partial
x}\right)^{-1}
\left({\partial^3 S_0 \over \partial x^3}\right)
\right]+
{1 \over 2m_0c^2}\left[m_0^2c^4
-(E-V)^2\right]=0\; ,
\end{eqnarray}
where $V$ is the potential, $E$ is the total energy of
the particle
of mass equal to $m_0$ at rest and $c$ is the
light velocity in
vacuum. The solution of Eq. (8) can be expressed by
Eq. (4), where
$\theta_1$ and $\theta_2$ represent now two real
independent
solutions of the Klein-Gordon equation
\begin{equation}
-c^2 \hbar^2 {\partial^2 \phi \over \partial x^2}+
\left[m_0^2c^4-(E-V)^2\right]\phi(x)=0\; .
\end{equation}

Taking advantage on these results, we will introduce in
the
following sections a relativistic quantum formalism
with which 
we can study the dynamics of high energy particles. In
Sec. 2, 
we present the relativistic quantum formalism with
which we 
establish, in Sec. 3, the expression of the conjugate
momentum.
Still in Sec. 3, we derive the relativistic quantum
law of motion.
Finally, we investigate in Sec. 4 the free particle
case and 
photon's trajectories. 

%
\vskip0.5\baselineskip
\noindent
{\bf 2\ \ Construction of the Relativistic Quantum
Lagrangian}
\vskip0.5\baselineskip
%

Before introducing the relativistic quantum Lagrangian,
let us recall
the Lagrangian formalism of special relativity. The
relativistic law
of motion can be derived from a Lagrangian of the form
\cite{SR}
\begin{equation}
L=-m_0c^2 \sqrt{1-{\dot{x}^2 \over c^2}}-V(x)\; ,
\end{equation}
where $c$ is the light velocity in vacuum, $V(x)$ is
the potential 
and $m_0$ the mass of the particle at rest. The
action of the
system is given by
\begin{equation}
S(x,t)=\int \ L\ dt\; ,
\end{equation}

Using expression (10) of the Lagrangian in the
Lagrange equation 
derived from the least action principle, we get to
\begin{equation}
{m_0\ddot{x} \over (1-\dot{x}^2/c^2)^{3 \over 2}}+{dV
\over dx}=0\; ,
\end{equation}
which represents the relativistic Newton's law.
Integrating
Eq. (12), we find
\begin{equation}
{m_0c^2 \over \sqrt{1-\dot{x}^2/c^2}}+V(x)=E\; ,
\end{equation}
where $E$ define the total energy of the particle
including 
the energy at rest $m_0c^2$. Eq. (14) represents
the 
relativistic conservation equation. We note that in
this case
the light velocity in vacuum $c$ appears as an upper
limit of 
the particle velocity since $\dot{x}$ and $c$ take
only real values.

Now consider a relativistic quantum system. As we have
noticed
for the quantum systems \cite{BD1} , the relativistic
quantum
reduced action $S_0$ expressed by Eq. (4) contains two
constants more than the usual constant $E$ appearing
in
the expression of the classical reduced action. This
suggests that 
the relativistic quantum law of motion is a fourth
order differential 
equation. Then, as in the quantum case \cite{BD1}, we
introduce in the
expression of the Lagrangian a function $f$ of $x$
depending on
a set $\Gamma$ of constants
playing the role of hidden parameters. As the
relativistic quantum 
Lagrangian must goes at the classical limit $(\hbar
\to 0)$ to 
the relativistic one, we postulate the following form
for the Lagrangian  
\begin{equation}
L=-m_0c^2 \sqrt{1-{\dot{x}^2 \over c^2}\;
f(x,\Gamma)}-V(x)\; ,
\end{equation}
in which the function $f(x,\Gamma)$ satisfies
\begin{equation}
 \lim_{\hbar \to 0} f(x,\Gamma)=1\; .
\end{equation}
The Hamiltonian corresponding to the Lagrangian (14)
can be
expressed by 
\begin{equation}
 H=P\dot{x}-L\; ,
\end{equation}
since $L$ depends only on the variables $x$, $\dot{x}$
and the 
set $\Gamma$ of constants, while the conjugate
momentum is
given by the relation
\begin{equation}
 P={\partial L \over \partial \dot{x}}=
{m_0\dot{x}f(x,\Gamma) \over
\sqrt{1-(\dot{x}^2/c^2)\; f(x,\Gamma)}}\; .
\end{equation}
At the classical limit, we see clearly by using Eq.
(15) that the
relativistic quantum momentum tends to the relativistic
one
expressed as
$$
 P={m_0\dot{x} \over 
\sqrt{1-²\dot{x}^2/c^2}}\; .
$$

Replacing Eqs. (17) and (14) in Eq. (16), one obtains 
\begin{equation}
H=\sqrt{m_{0}^2c^4+{P^2c^2 \over f}}+V(x)\; ,
\end{equation}
For the stationary cases the Hamiltonian $H$
corresponds to the total
energy $E$ of the particle. Then, we can write Eq.
(19) as (after taking in account Eq. (18))
\begin{equation}
E={m_0c^2 \over
\sqrt{1-(\dot{x}^2/c^2)f(x,\Gamma)}}+V(x)\; ,
\end{equation}
This last relation can be deduced from the Lagrangian
expressed
by (15). Indeed, the least action principle leads to
\begin{equation}
{m_0\ddot{x}f \over
\sqrt{1-(\dot{x}^2/c^2)\; f}}+{m_0 \over 2}\dot{x}\;
{df/dt \over \sqrt{1-(\dot{x}^2/c^2)\;
f}}+m_0\dot{x}f{d \over dt}
\left({1 \over \sqrt{1-(\dot{x}^2/c^2)\;
f}}\right)+{dV \over dx}=0\; ,
\end{equation}
where we have used the fact that $\dot{x}(\partial f /
\partial x)=df/dt$.
After integrating Eq. (21), one finds Eq. (20).

Now, let us establish the explicit form of $f$. From
(20), we can write
\begin{equation}
{1 \over 2}{m_0\dot{x}^2 f \over
1-(\dot{x}^2/c^2)\; f}+{1 \over 2m_0c^2}\left[m_0^2c^4
-(E-V)^2\right]=0\; .
\end{equation}
Using Eq. (18) in Eq. (22), we obtain
\begin{equation}
{P^2 \over 2m_0}{1 \over f(x,\Gamma)}+{1 \over
2m_0c^2}\left[m_0^2c^4
-(E-V)^2\right]=0\; ,
\end{equation}
from which and taking into account Eq. (2), we deduce
\begin{equation}
f(x,\Gamma)={c^2\; ({\partial S_0 / \partial x})^2
\over \left[m_0^2c^4
-(E-V)^2\right]}\; .
\end{equation}
If we substitute $S_0$ with its expression (4) in Eq.
(24), we notice
that $f$ depends on $x$, $a$, $b$ and $E$. Then, we
identify the 
set $\Gamma$ to the integration constants $a$, $b$ and
$E$.
Remark that at the classical limit $(\hbar \to 0)$, we
have from the
RQSHJE (Eq. (9))
$$
({\partial S_0 / \partial x})^2={1 \over c^2}
\left[(E-V)^2-m_0^2c^4\right]\; ,
$$
and then, we obtain in Eq. (24)
$$
\lim_{\hbar \to 0} f(x,\Gamma)=1\; ,
$$
which corresponds to Eq. (16). From Eq. (24), we note
that the
function $f(x,a,b)$ is real positive when $E-V>m_0c^2$
(classically allowed regions) and real negative when
$E-V<m_0c^2$
(classically forbidden regions). We note also From Eq.
(20) that
$c$ does not appear as an upper limit since the square
root
contains the function $f$depending on $x$, while for
the classically
forbidden regions there is no limit of the particle
velocity since $f$
is negative.

Now, as it is the case for the quantum Lagrangian
\cite{BD1}, let us
demonstrate that the use of expressions (15) for the
Lagrangian
and (19) for the Hamiltonian is justified by appealing
to a coordinate
transformation $x \to \hat{x}$ after which, the RQSHJE
takes the form
\begin{equation}
{1 \over 2m_0}\left({\partial \hat{S}_0(\hat{x}) \over
\partial \hat{x}}\right)^2+{1 \over 2m_0c^2} 
\left[m_0^2c^4-(E-\hat{V}(\hat{x}))^2\right]=0
\end{equation}
The coordinate $\hat{x}$ defined by
\begin{equation}
\hat{x}=\int^{x} {c\; ({\partial S_0 / \partial x})
\over
\sqrt{(E-V)^2-m_0^2c^4}}\; dx\; ,
\end{equation}
and which we will call ``{\it Relativistic quantum
coordinate}''
is analogue to the quantum coordinate introduced by
Faraggi 
and Matone in Ref. \cite{FM3,FM5}.

By setting
\begin{equation}
\hat{S}_0(\hat{x})= S_0(x)\; \; \; ,
\hat{V}(\hat{x})=V(x)\; ,
\end{equation}
Eq. (25) takes the form
\begin{equation}
{1 \over 2m_0}\left({\partial S_0 \over 
\partial x}\right)^2\left({\partial x \over
\partial \hat{x}}\right)^2+{1 \over 2m_0c^2} 
\left[m_0^2c^4-(E-V(x))^2\right]=0\; .
\end{equation}
Eq. (28) is equivalent to Eq. (20) since, as we can
deduce from Eqs.
(24) and (26)
\begin{equation}
f(x,a,b,E)=\left({\partial x \over 
\partial \hat{x}}\right)^2\; .
\end{equation}
Then, we note that a classical formulation with
respect to the 
relativistic quantum coordinate $\hat{x}$ is strictly
equivalent 
to both Lagrangian and Hamiltonian formulations
(Eqs. (15) and (19)).

%
\vskip0.5\baselineskip
\noindent
{\bf 3\ \ The Relativistic Quantum Law of Motion}
\vskip0.5\baselineskip
%
 
Now, taking Eq. (24) into Eq. (20), we get  to the
following
expression of the conjugate momentum
\begin{equation}
{\partial S_0 \over \partial x}={E-V(x) \over
\dot{x}}-
{m_0^2 c^4 \over (E-V)\dot{x}} ,
\end{equation}
where we have eliminated one of the roots for
$\partial S_0/\partial x$
since Eq. (18) indicates that $\dot{x}$ and
$P=\partial S_0/\partial x$
have the same sign in classically allowed regions and
are opposites
in forbidden ones. Note that it is also possible to
derive Eq. (30) by
using a relativistic quantum version of the Jacobi's
theorem, exactly
in the same way as it is done in Ref. \cite{BD1}.

Let us now consider both relativistic and classical
limits in Eq. (30).
For the relativistic limit $(c \to \infty)$, the
kinetic energy $T$
given by
\begin{equation}
T=E-V-m_0c^2\; ,
\end{equation}
satisfies the relation
\begin{equation}
T<<m_0c^2\; .
\end{equation}
Using Eq. (31) in Eq. (31), we have
\begin{equation}
\dot{x}{\partial S_0 \over \partial x}=
T+m_0c^2-{m_0^2c^4 \over T+m_0c^2}\; ,
\end{equation}
which reduces at the relativistic limit to
\begin{equation}
{\partial S_0 \over \partial x}={2\; T \over
\dot{x}}\; ,
\end{equation}
after having used the fact that
$(1+T/m_0c^2)^{-1}\simeq(1-T/m_0c^2)$.
Eq. (34) corresponds to the quantum expression of the
conjugate
momentum given in Ref. \cite{BD1}  as
\begin{equation}
{\partial S_0 \over \partial x}=2\; {(E^{'}-V) \over
\dot{x}}\; ,
\end{equation}
where $E^{'}-V$ represents the kinetic energy for the
non-relativistic cases.

\noindent
For the classical limit $(\hbar \to 0)$, we have$ f
\to 1$,
then Eq. (30) reduces to
\begin{equation}
\dot{x}{\partial S_0 \over \partial x}=
{m_0c^2 \over \sqrt{1-\dot{x}^2/c^2}}-
m_0c^2 \sqrt{1-\dot{x}^2/c^2}\; ,
\end{equation}
from which we deduce
\begin{equation}
{\partial S_0 \over \partial x}=
{m_0\dot{x} \over \sqrt{1-\dot{x}^2/c^2}}\; ,
\end{equation}
representing the expression of the conjugate momentum
for a
relativistic system. Thus, at the classical limit
$(\hbar \to 0)$
the relativistic quantum motion reduces to the
relativistic one,
and, at the relativistic limit $(c \to \infty)$, it
reduces to the
quantum one. Then, we can consider expression (30) as
a
generalization of the conjugate momentum to the
relativistic
quantum systems.

Now, we will derive the relativistic quantum equation
of motion
using expression (30) of the conjugate momentum. Then,
if one computes the derivatives
\begin{equation}
{\partial^2 S_0 \over \partial x^2}=-{1 \over
\dot{x}}{dV \over dx}
\left[1+{m_0^2c^4 \over (E-V)^2}\right]+
{\ddot{x} \over \dot{x}^3}{1 \over E-V}
\left[m_0^2c^4-(E-V)^2 \right]\; ,
\end{equation}
\begin{eqnarray}
\hskip-7mm {\partial^3 S_0 \over \partial x^3}=
-{1 \over \dot{x}}{d^2V \over dx^2}
\left[1+{m_0^2c^4 \over (E-V)^2}\right]+
\left(3{\ddot{x}^2 \over \dot{x}^5}-{\dot{\ddot{x}}
\over \dot{x}^4}\right)
\left[{(E-V)^2-m_0^2c^4 \over
(E-V)}\right]+\hskip-17mm&& \nonumber\\
2{\ddot{x} \over \dot{x}^3}{dV \over dx}
\left[{(E-V)^2+m_0^2c^4 \over (E-V)^2}\right]-
{2 \over \dot{x}}\left({dV \over dx}\right)^2{m_0^2c^4
\over (E-V)^3} \; ,
\end{eqnarray}
one can deduce the following expression of the
Schwarzian
derivative of $S_0$ with respect to $x$.
\begin{eqnarray}
\{S_0,x\}=\left({\dot{\ddot{x}} \over \dot{x}^3}-{3
\over 2}
{\ddot{x}^2 \over \dot{x}^4}\right)+
\left({\ddot{x} \over \dot{x}^2}{dV \over dx}+{d^2V
\over dx^2}\right)
{(E-V)^2+m_0^2c^4 \over
(E-V)^2-m_0^2c^4}+\hskip-18mm&& \nonumber\\
{3 \over 2}{1 \over (E-V)^2}\left(dV \over dx\right)^2
\left[{(E-V)^2+m_0^2c^4 \over
(E-V)^2-m_0^2c^4}\right]^2+
\hskip-2mm&& \nonumber\\
{2 \over(E-V)^2 }\left({dV \over dx}\right)^2{m_0^2c^4
\over (E-V)^2-m_0^2c^4}
=0\; ,
\end{eqnarray}
Replacing Eqs. (30) and (40) in Eq. (9), we get to
\begin{eqnarray}
\left[(E-V)^2-m_0^2c^4\right]^2+{\dot{x}^2 \over
c^2}(E-V)^2
\left[(E-V)^2-m_0^2c^4\right]+{\hbar^2 \over 2}
\left[{3 \over 2}\left({\ddot{x} \over
\dot{x}}\right)^2-
{\dot{\ddot{x}} \over \dot{x}}\right] \cdot
\hskip-10mm&& \nonumber\\
(E-V)^2-{\hbar^2 \over 2}\left(\ddot{x}{dV \over dx}+
\dot{x}^2{d^2V \over dx^2}\right)
\left[{(E-V)^2+m_0^2c^4 \over (E-V)^2-m_0^2c^4}\right]
(E-V)^2-
{3\hbar^2 \over 4}\cdot
\hskip-1mm&& \nonumber\\
\left(\dot{x}{dV \over dx}\right)^2
\left[{(E-V)^2+m_0^2c^4 \over
(E-V)^2-m_0^2c^4}\right]^2-
\hbar^2\left(\dot{x}{dV \over dx}\right)^2{m_0^2c^4
\over (E-V)^2-m_0^2c^4}
=0\; .
\end{eqnarray}
Because it depends on the integration constant $E$,
this equation
represents a first integral of the relativistic
quantum Newton's
law (FIRQNL). As we observe, it is a third order
differential
equation in $x$ containing the first and the second
derivatives
of the potential $V$ with respect to $x$. Then, the
solution
$x(t,E,a,b,c)$ of (41) contains four integration
constants which
can be determined by the initial conditions.
It is clear that if we set $\hbar=0$, Eq. (41) reduces
to Eq.(14)
representing the relativistic conservation equation.
Note also that,
after taking the relativistic limit $(c \to \infty)$
in Eq. (41),
one gets to Eq. (7) representing the FIQNL.

Remark that the FIRQNL can be derived by using the
solution (4)
of the RQSHJE as the same way presented in Ref.
\cite{BD1} to derive
the FIQNL. This method leads straightforwardly to Eq.
(41).

 %
\vskip0.5\baselineskip
\noindent
{\bf 4\ \ The Free Particle Case and Photon's
trajectories}
\vskip0.5\baselineskip
%

We will study now the motion of a free particle with
$m_0$ rest mass. For this case the expressions (30)
of the conjugate
momentum and (41) of the FIRQNL reduce respectively to
\begin{equation}
{\partial S_0 \over \partial x}={E \over
\dot{x}}-{m_0^2c^4 \over \dot{x}E}\; ,
\end{equation}
and
\begin{equation}
(E^2-m_0^2c^4)^2-{\dot{x}^2 \over c^2}\; E^2 \;
(E^2-m_0^2c^4)+
{\hbar^2 \over 2}\left[{3 \over 2}\left({\ddot{x}
\over \dot{x}}\right)^2-
{\dot{\ddot{x}} \over \dot{x}}\right]E^2=0\; .
\end{equation}
In order to determine $x(t,E,a,b,c)$, we can solve Eq.
(43) or
Eq. (42) with appealing to the solutions of the
Klein-Gordon equation
for a vanishing potential. Indeed, in this purpose let
us introduce
the new variables
\begin{equation}
U={1 \over c}\sqrt{E^2-m_0^2c^4}\; \ ; \;
q={c \over E}\sqrt{E^2-m_0^2c^4}\; t\; ,
\end{equation}
with which Eq. (43) takes the form
\begin{eqnarray}
{1 \over 2m_0} \left({dU \over dq}\right)^2-{\hbar^2
\over 4m_0}
\left[{3 \over 2}\left({dU \over dq}\right)^{-2}
\left({d^2U \over dq^2}\right)^2-\left({dU \over
dq}\right)^{-1}
\left({d^3U \over dq^3}\right)\right]+\hskip-17mm&&
\nonumber\\
{1 \over 2m_0}[m_0^2c^4-E^2]=0\; ,
\end{eqnarray}
Note that $U$ and $q$ have the dimensions of,
respectively,
action and distance. Eq. (45) has as solution
\begin{equation}
U={\hbar} \arctan{\left[a {{\psi_1}\over{\psi_2}}+b
\right]}
+ U_0 \; ,
\end{equation}
$a$ and $b$ being real constants, $\psi_1$ and
$\psi_2$ are two
solutions of the Klein-Gordon equation with vanishing
potential
\begin{equation}
-c^2 \hbar^2 {\partial^2 \psi \over \partial x^2}+
\left[m_0^2c^4-E^2\right]\psi(x)=0\; .
\end{equation}
If we choose the two solutions of Eq. (47) as
$
{\psi_1}=\sin{\left(\sqrt{E^2-m_0^2c^4}\; q /
{\hbar}c\right)}\
$
and
$
\ {\psi_2}=\cos{\left(\sqrt{E^2-m_0^2c^4}\; q /
{\hbar}c\right)},
$
Eq. (46) will be written as
\begin {equation}
x(t)={\hbar c \over \sqrt{E^2-m_0^2c^4}}
\arctan{\left[a \tan{\left({E^2-m_0^2c^4
\over\hbar E}\; t\right)}+b\right]}+x_0  \; .
\end {equation}
This equation represents the time equation of
trajectories.
As we have mentioned above, $x(t)$ contains four
constants
since the fundamental equation of motion is a fourth
order
differential equation. Because the arctangent function
is
defined on the interval $\rbrack-\pi/2,\pi/2 \;
\lbrack$,
Eq. (48) shows that the particle is contained between
$$
-{\hbar c  \over \sqrt{E^2-m_0^2c^4}} {\pi \over
2}+x_0
$$
and
$$
{\hbar c  \over \sqrt{E^2-m_0^2c^4}} {\pi \over 2}+x_0
\; .
$$
This is not possible since the particle is free. Then,
it is
necessary to readjust the additive integration
constant $x_0$
after every interval of time in which the tangent
function goes
from $-\infty$ to $+\infty$ in such a way to guarantee
the
continuity of $x(t)$. In this purpose, expression (48)
must be
rewritten as
\begin {equation}
x(t)={\hbar c \over \sqrt{E^2-m_0^2c^4}}
\arctan{\left[a \tan{\left({E^2-m_0^2c^4
\over\hbar E}\; t\right)}+b\right]}+{\pi \hbar c \over
\sqrt{E^2-m_0^2c^4}}n+x_0  \; .
\end {equation}
with
$$
t\; \in \left[{\pi \hbar E  \over
E^2-m_0^2c^4}\left(n-{1 \over 2}\right)
; {\pi \hbar E  \over E^2-m_0^2c^4}\left(n+{1 \over
2}\right) \right]
$$
for every integer number. The purely relativistic
trajectory is obtained
for $a=1$ and $b=0$. Indeed, for these values, Eq.
(49) reduces
to the relativistic relation
\begin {equation}
x(t)= {c \over E}\; \sqrt{E^2-m_0^2c^4}\; t+x_0  \; .
\end {equation}
Eq. (49) indicates that all the trajectories defined
by the
set $(a,b)$ pass through some nodes exactly
as we have seen for quantum trajectories \cite{BD2}.
These nodes correspond to the times
\begin {equation}
t_n={\pi \hbar E  \over E^2-m_0^2c^4}\left(n+{1 \over
2}\right)\; .
\end {equation}
The distance between two adjacent nodes is on time
axis
\begin {equation}
\Delta t_n=t_{n+1}-t_n={\pi \hbar E  \over
E^2-m_0^2c^4}\; .
\end {equation}
and space axis
\begin {equation}
\Delta x_n=x_{n+1}-x_n={\pi \hbar c  \over
\sqrt{E^2-m_0^2c^4}}\; .
\end {equation}
These distances are both proportional to $\hbar$
meaning that at
the classical limit $\hbar \to 0$ the nodes becomes
infinitely
near, and then, all quantum trajectories tend to the
purely
relativistic one. As it is explained in Ref.
\cite{BD2},
this is the reason why in problems for which $\hbar$
can
be disregarded, relativistic quantum trajectories
reduces to the
purely relativistic one.

Now, let us investigate the case where the free
particle is a photon
which have a vanishing mass at rest $(m_0=0)$.
Thus, for
the photon, the conjugate momentum is
\begin {equation}
{\partial S_0 \over \partial x}= {E \over \dot{x}}\; .
\end {equation}
and the FIRQNL is
\begin {equation}
E^2-\dot{x}^2{E^2 \over c^2}+
{\hbar^2 \over 2}
\left[{3 \over 2}\left({\ddot{x} \over
\dot{x}}\right)^2-
{\dot{\ddot{x}} \over \dot{x}}\right]=0\; .
\end {equation}
From Eqs. (54) and (55), we can establish the time
equation of photon's
trajectories
\begin {equation}
x(t)={\hbar c \over E}
\arctan{\left[a \tan{\left({E \over \hbar }\;
t\right)}+
b\right]}+{\pi \hbar c \over E}\; n+x_0  \; .
\end {equation}
with
$$
t \in \left[{\pi \hbar \over E}(n-{1 \over 2}),
{\pi \hbar \over E}(n+{1 \over 2})\right]
$$
We can check easily that the purely relativistic
trajectory can be
obtained from (56) by setting $a=1$ and $b=0$, we then
find
\begin {equation}
x(t)=ct+x_0\; .
\end {equation}
As the free massive particle case, we note
that all these trajectories even the purely
relativistic one
$(a=1, b=0)$ pass through nodes corresponding to
the times
\begin {equation}
t_n={\pi \hbar \over E}\left(n+{1 \over 2}\right)\; ,
\end {equation}
for which the positions
\begin {equation}
x_n={\pi \hbar c \over E}\left(n+{1 \over 2}\right)\;
,
\end {equation}
do not depend on $a$ and $b$. The distance between
two adjacent
nodes on time axis
\begin {equation}
\Delta t_n=t_{n+1}-t_n={\pi \hbar \over E}\; ,
\end {equation}
and space axis
\begin {equation}
\Delta x_n=x_{n+1}-x_n={\pi \hbar c \over E}\; ,
\end {equation}
are both proportional to $\hbar$, meaning that at the
classical limit
$(\hbar \to 0)$, the nodes become infinitely near, and
all relativistic
quantum trajectories of the photon tend to the purely
relativistic one.
In order to check this fact, we refer the reader to
Ref. \cite{BD2}.

An important remark should be made concerning
the fact that the particle's velocity is less
than $c$ in some regions and great than $c$ in
other regions. This fact contradict the
noticing, in special relativity that the light
velocity in
vacuum $c$ is the upper limit of the velocities of any
massive
particle. However, we note from
Eqs. (49), (52), (53), (56), (60) and (61) that the
free particle and the
photon cover the distance
between two nodes in such a way that the average
velocity between
two nodes is
\begin {equation}
v_{mean_{el}}={\Delta x_n \over\Delta t_n}={c
\sqrt{E^2-m_0^2c^4}
\over E}\; ,
\end {equation}
for the electron, and
\begin {equation}
v_{mean_{ph}}={\Delta x_n \over\Delta t_n} = c \; ,
\end {equation}
for the photon. The velocity given by Eq. (62) is the
purely
relativistic one (see Eq. (50)), it is less than $c$,
while the
velocity given by Eq. (63) is the light velocity in
vacuum $c$.
These results can be interpreted as follows:
In subquantic scales, the light velocity does not
appear as an
upper limit, otherwise, the particle may have a velocity greater
than $c$
between two nodes. But, in the average and
between
two or more than two nodes, $c$ appears as an upper
limit,
since for all possible trajectories the distance
between two
nodes (Eqs. (53) and (61)) is covered  at constant
times
(Eqs. (52) and (60)) making that the average velocity
is
less than (Eq. (62)) or equal (Eq. (63)) to $c$.

Let us now consider the motion in classically
forbidden
regions $(E<m_0c^2)$. For this case, the solutions of
Klein-Gordon equation are
\begin {equation}
\psi_1=exp{\left(-{\sqrt{m_0^2c^4-E^2} \over \hbar \;
c}\; x\right)}\; ,
\end {equation}
and
\begin {equation}
\psi_2=exp{\left(+{\sqrt{m_0^2c^4-E^2} \over \hbar \;
c}\; x\right)}\; .
\end {equation}
Then, by integrating Eq. (42), we get
\begin{equation}
S_0=\left( E-{m_0^2c^4 \over E} \right)\; t,
\end{equation}
Replacing Eqs. (64) and (65) in Eq. (66), and taking
in account
expression (4) of $S_0$, we find
\begin{equation}
x(t) = {\hbar \; c \over 2\sqrt{m_0^2c^4-E^2}}\ln
{\left\vert
a \tan \left({E^2-m_0^2c^4
\over\hbar E}\; t\right)+b\right\vert}
+ x_0 \; .
\end{equation}
This equation represents the relativistic quantum time
equation for a particle moving in the classically
forbidden
regions. Its velocity is given by
\begin {equation}
\dot{x}(t) = -{a\; c \over 2\; E}  \;
{1+ \tan^2 ({E^2-m_0^2c^4 / \hbar E}) \over
a \tan ({E^2-m_0^2c^4 / \hbar E}) +b } \; ,
\end {equation}
from which we can see that at the times
$-(2n+1)\pi \hbar E /4(E^2-m_0^2c^4)$ the velocity
becomes
infinite. Let us recall that we have already faced
in Ref. \cite{BD2} infinite velocities for quantum
problems.
Nevertheless, in non-relativistic problems such
velocities do not
contradict any postulate, while for the relativistic
problems the
infinite velocities appears to contradict the fact
that in special
relativity the light velocity in vacuum $c$ is an
upper limit.

To sustain our finding, let us recall the results of
R. Y. Chiao's experiments \cite{Chiao}. Chiao asserts
that
{\it " Experiments have shown that individual photons
penetrate
an optical tunnel barrier with an effective group
velocity
considerably greater than the vacuum speed of light " }.
The experiments were done with photon pairs emitted in
slightly different directions so that one photon
passed through
the tunnel barrier, while the other photon passed
through the vacuum. Chiao found that the photon's
transit time
through the barrier was smaller than the twin photon's
transit
time through an equal distance in vacuum. These
results
are in agreement with our approach, since it
illustrate infinite
(greater than $c$) velocities in classically
forbidden regions (tunnel barrier).

{\bf Conclusions}
To conclude, we would like to stress that  we exposed
in
this article an original approach of the relativistic
quantum
mechanics. It is a generalization of the one exposed
in Ref. \cite{BD1}. Thus, we have derived the
relativistic quantum Newton's law (41) from (30) which
represents the relation between the conjugate momentum
and the speed of the particle. These two
equations  are obtained in different contexts:
\begin{itemize}
\item a Lagrangian formulation;

\item a Hamiltonian formulation,
 grounding oneself on the solution of the RQSHJE;

\item a relativistic quantum version of Jacobi's theorem

\end{itemize}

\vskip\baselineskip

\noindent
{\bf ACKNOWLEDGEMNTS}

I would like to think Dr. F. Djama for encouragement and help.

\noindent
\newpage

%
\vskip\baselineskip
\noindent
{\bf \ REFERENCES}
\vskip\baselineskip
%


\begin{enumerate}

\bibitem {FM1}
A.~E.~Faraggi and M.~Matone,  {\it Phys. Lett.} B 450,
34 (1999); hep-th/9705108.

\bibitem{FM2}
A.~E.~Faraggi and M.~Matone,  {\it Phys. Lett.} B 437,
369 (1998); hep-th/9711028.

\bibitem{FM3}
A.~E.~Faraggi and M.~Matone, {\it Int. J. Mod. Phys.}
A 15, 1869
(2000); hep-th/9809127.

\bibitem{FM4}
G.~Bertoldi, A.~E.~Faraggi and M.~Matone,  {\it Class.
Quant. Grav.} 17,
3965 (2000); hep-ph/9909201.

\bibitem{Floyd1}
 E.~R.~Floyd, {\it Phys. Rev.} D 34, 3246 (1986).

\bibitem{Floyd2}
E.~R.~Floyd, {\it Found. Phys. Rev.} 9, 489(1996);
quant-ph/9707051.

\bibitem{Floyd3}
E.~R.~Floyd, {\it Phys. Lett.} A 214, 259 (1996);

\bibitem{Floyd4}
 E.~R.~Floyd,  quant-ph/0009070.

\bibitem{BD1}
A. Bouda and T. Djama, {\it Phys. Lett.} A 285, 27
(2001); quant-ph/0103071.


\bibitem{BD2}
A. Bouda and T. Djama, quant-ph/0108022.

\bibitem{SR}
Boudenot, Electromagnétisme et gravitation
relativiste.

\bibitem{FM5}
A. E. Faraggi and M. Matone, {\it Phys. Lett.} A
249,180 (1998);
hep-ph/9801033.

\bibitem{Chiao}
R.~Y.~Chiao, quant-ph/9811019.


\end{enumerate}
\end{document}